\documentclass[12pt]{article}

\usepackage[letterpaper,hmargin=1in,vmargin=1in]{geometry}

\usepackage{graphicx,epstopdf,amsmath,amsfonts}

\def\be{\begin{equation}}
\def\ee{\end{equation}}
\def\ba{\begin{eqnarray}}
\def\ea{\end{eqnarray}}
\def\ge{\mathrel{\raise.3ex\hbox{$>$\kern-.75em\lower1ex\hbox{$\sim$}}}}
\def\la{\mathrel{\raise.3ex\hbox{$<$\kern-.75em\lower1ex\hbox{$\sim$}}}}

\def\simgt{\mathrel{\raise.3ex\hbox{$>$\kern-.75em\lower1ex\hbox{$\sim$}}}}
\def\simlt{\mathrel{\raise.3ex\hbox{$<$\kern-.75em\lower1ex\hbox{$\sim$}}}}

\newcommand{\bi}[1]{\bibitem{#1}}
\newcommand{\fr}[2]{\frac{#1}{#2}}

\newcommand{\nc}{\newcommand}

\nc{\gone}{\bar g_{\pi NN}^{(1)}}
\nc{\gzero}{\bar g_{\pi NN}^{(0)}}
\nc{\al}{\alpha}
\nc{\ga}{\gamma}
\nc{\de}{\delta}
\nc{\ep}{\epsilon}
\nc{\ze}{\zeta}
\nc{\et}{\eta}
\nc{\ka}{\kappa}
\nc{\rh}{\rho}
\nc{\si}{\sigma}
\nc{\ta}{\tau}
\nc{\up}{\upsilon}
\nc{\ph}{\phi}
\nc{\ch}{\chi}
\nc{\ps}{\psi}
\nc{\om}{\omega}
\nc{\Ga}{\Gamma}
\nc{\De}{\Delta}
\nc{\La}{\Lambda}
\nc{\Si}{\Sigma}
\nc{\Up}{\Upsilon}
\nc{\Ph}{\Phi}
\nc{\Ps}{\Psi}
\nc{\Om}{\Omega}
\nc{\ptl}{\partial}
\nc{\del}{\nabla}
\nc{\ov}{\overline}
\nc{\newcaption}[1]{\centerline{\parbox{15cm}{\caption{#1}}}}

\def\beq{\begin{equation}}
\def\eeq{\end{equation}}
\def\bmat{\begin{displaymath}}
\def\emat{\end{displaymath}}
\def\bear{\begin{eqnarray}}
\def\eear{\end{eqnarray}}
\def\ba{\begin{eqnarray}}
\def\ea{\end{eqnarray}}
\def\bery{\begin{array}}
\def\ery{\end{array}}
\def\bit{\begin{itemize}}
\def\eit{\end{itemize}}
\def\ben{\begin{enumerate}}
\def\een{\end{enumerate}}
\def\btab{\begin{tabular}}
\def\etab{\end{tabular}}
\def\btbl{\begin{table}}
\def\etbl{\end{table}}
\def\bfig{\begin{figure}[htb]}
\def\efig{\end{figure}}
\def\bpic{\begin{picture}}
\def\epic{\end{picture}}

\def\ga{\mathrel{\raise.3ex\hbox{$>$\kern-.75em\lower1ex\hbox{$\sim$}}}}
\def\la{\mathrel{\raise.3ex\hbox{$<$\kern-.75em\lower1ex\hbox{$\sim$}}}}
\def\gappeq{\mathrel{\rlap {\raise.5ex\hbox{$>$}}
{\lower.5ex\hbox{$\sim$}}}}
\def\lappeq{\mathrel{\rlap{\raise.5ex\hbox{$<$}}
{\lower.5ex\hbox{$\sim$}}}}

\def\gyr{{\rm \, G\kern-0.125em yr}}
\def\mev{{\rm \, Me\kern-0.125em V}}
\def\gev{{\rm \, Ge\kern-0.125em V}}
\def\tev{{\rm \, Te\kern-0.125em V}}

\begin{document}

\begin{titlepage}

\setcounter{page}{1}

\vspace*{0.3in}

\begin{center}

{\Large \bf \begin{center}
Astrophysical Signatures of Secluded Dark Matter\\
\end{center}}

\vspace*{1cm}
\normalsize

{\bf  Maxim Pospelov$^{\,(a,b)}$ and Adam Ritz$^{\,(a)}$ }

\smallskip
\medskip

$^{\,(a)}${\it Department of Physics and Astronomy, University of Victoria, \\
     Victoria, BC, V8P 1A1 Canada}

$^{\,(b)}${\it Perimeter Institute for Theoretical Physics, Waterloo,
ON, N2J 2W9, Canada}

\smallskip
\end{center}
\vskip0.2in

\centerline{\large\bf Abstract}

We analyze the indirect astrophysical signatures of secluded models of WIMP dark matter, characterized
by a weak-scale rate for annihilation into light MeV-scale mediators which are metastable to decay into Standard Model
states. Such scenarios allow a significant enhancement of the annihilation cross section in the galactic halo relative to its value at
freeze-out, particularly when the mediator is light enough for this process to proceed through
radiative capture to a metastable `WIMP-onium' bound state. For MeV-scale vector mediators charged under a hidden
U(1)$'$ gauge group, the enhanced annihilation rate leads predominantly to a sizable excess positron
flux, even in the absence of astrophysical boost factors.

\vfil
\leftline{October 2008}

\end{titlepage}

\subsection*{1. Introduction}

The growing level of astrophysical and cosmological evidence for dark matter has led in recent years to
an expansion in experimental programs that aim to detect its non-gravitational interactions \cite{Fuller:2007hk}. These
range from direct production at colliders to the recoil of 
galactic dark matter on nuclei in underground detectors, and indirect detection 
of annihilation products from various parts of the galactic halo. The primary theoretical motivation for anticipating interactions
with the Standard Model (SM) at a detectable level is the apparent simplicity of the thermal WIMP paradigm. Namely,
the observation that a thermal relic with weak scale mass and annihilation cross-section  naturally provides roughly the correct 
cosmological abundance \cite{LW}. However, even this additional theoretical input still leaves a vast range of
possibilities, and thus it is far from clear which of the various experimental or observational strategies will have the
best sensitivity. It is therefore important to ensure that the observational reach is as broad as possible, given reasonable
theoretical expectations, and to that end it is useful to explore the differing levels of sensitivity for various detection
strategies within the simplest consistent models for thermal dark matter.

A generic problem is that while the thermal WIMP scenario imposes constraints on the annihilation cross-section,
it clearly does not directly constrain the scattering cross-section with matter, and it is possible to consider
scenarios where these rates differ significantly. An example is the class of {\it secluded} WIMPs \cite{prv1}, or 
alternatively a dark hidden sector, where WIMPs annihilate with a weak-scale rate to metastable mediators which are 
in turn very weakly coupled to the Standard Model. In such scenarios, the direct scattering cross-section is negligible,
and the primary detection strategy is via observation of the products of annihilation in parts of the galactic halo.
Various annihilation products may be observed, i.e. $\gamma$-rays, anti-matter (positrons and anti-protons), and also
neutrinos from annihilation in the sun and elsewhere, 
depending on the form of the model. Thus, it is important
to know which if any of these annihilation products would provide 
the best sensitivity given current and future observational capabilities. 
Such an investigation is timely given that existing data from the HEAT \cite{heat} 
and AMS-01 \cite{ams1} experiments indicating
a possible excess in the relative positron fraction $n_{e^+}/(n_{e^-}+n_{e^+})$, and 
thus a possible enhancement in the production of 
$O(10\,$GeV) positrons in the galactic vicinity of the solar system, recently received
support from preliminary results from PAMELA \cite{pamela}. Although such an excess - even if 
confirmed to be above the expected background level 
produced via cosmic rays \cite{background} -- could be due to various astrophysical sources, 
e.g. pulsars in the local neighbourhood, 
the recent announcement has generated a flurry of 
theoretical activity addressing the question
of whether a positron excess could be tied to WIMP annihilation  \cite{th-speculation}. 
This recent work has built upon several earlier analyses \cite{olderDM,DMcosmic}, and 
a fairly general conclusion
is that an excess at the level apparently observed by HEAT and AMS-01 (and PAMELA) 
would require a nontrivial enhancement of the 
annihilation rate in the galaxy relative to its rate at freeze-out for a thermally populated WIMP. 
Such enhancements are often 
conjectured (in the form of boost factors) to result from local features of the halo distribution, 
but more generally this raises
a challenge to see whether generic WIMP scenarios could produce an anti-matter flux at a level sufficiently far 
above background to be
observed.

In this note, in the context of secluded WIMP models, we point out that if the mass of the mediator $m_V$ 
satisfies the following bounds, 
\be
2 m_e \la m_V \la p_{\rm DM} \la m_{\rm DM} \al'
\label{condition1}
\ee
where $p_{\rm DM}=m_{\rm DM} v$ is the typical nonrelativistic  WIMP momentum inside the galactic halo, and 
$\alpha'$ is the strength of the interaction in the dark sector, the resulting annihilation rate is 
enhanced over the cosmological rate by a factor of $O(\alpha'/v)\ga 1$, while the decay products consist mostly of 
electron and muon pairs and photons. Most importantly, if $m_V$ is less than the `WIMP-onium' binding energy $E_{\rm bind}$, i.e. if
\be
4m_V < (\alpha')^2m_{\rm DM }  ,
\label{condition2}
\ee
then annihilation inside the halo proceeds via the intermediate formation of a 
metastable WIMP-onium bound state, and the enhancement factor over the 
cosmological annihilation rate grows to $O(10\,\alpha'/v_{\rm DM })$. Interestingly enough, this
appears to be of sufficient magnitude to boost the observed positron fraction in the multi-GeV range to
a level that would be observed as an excess above background without the need for astrophysical
boost factors.

To provide a concrete example, we calculate the 
indirect detection signal in a model of secluded dark matter
with a vector U(1)$'$ mediator $V$ kinetically mixed with the photon, noting that similar 
results can be obtained for scalar mediation. 
In the next section, we discuss the various annihilation processes, 
and the distinctions between those most relevant for comological 
freeze-out, and for producing observable signals in the galaxy today. In Section~3, we 
briefly discuss the observational
implications of galactic annihilation focusing on excess positrons and $\gamma$'s, and 
conclude in Section~4 with some other implications of this dark matter scenario.

\subsection*{2. Annihilation of Secluded WIMPs}

The characteristic feature of secluded WIMPs \cite{prv1} is that the WIMP in the hidden sector communicates with the 
Standard Model via  a mediator, with $m_{\rm WIMP}> m_{\rm mediator}$, such that direct annihilation into a pair of mediators is
always possible. The correct relic density may then be achieved while imposing only very mild constraints on the coupling of
the mediator to the SM  -- a mediator lifetime of up to a second, so that decays occur before primordial nucleosynthesis (BBN), is sufficient -- and this
allows the WIMP to be secluded from the SM with a very small cross-section for direct scattering. This generic feature leads us to
focus on the annilation process as a means for indirect detection of such dark matter scenarios. 

The annihilation details are sensitive to the relevant mass scales 
of the WIMP and the mediator, and scenarios in which the mediator may 
couple to the SM at the renormalizable level are of particular 
interest since the mediator can then naturally be rather light. There are only a limited number of options 
for introducing such couplings, often referred 
 to as portals, and most of them have been considered at length in a number of publications \cite{portal1,portal2}. 
One of the most natural possibilities that leaves the mass-scale of 
 the mediator as a free parameter is the kinetic mixing portal between the 
hypercharge gauge boson $B_\mu$ and a U(1)$'$ mediator $V_\mu$ \cite{holdom}. The only renormalizable coupling to the SM occurs through
 kinetic mixing, ${\cal L}_{\rm mix} \sim \ka V_{\mu\nu}B_{\mu\nu}$, and the
 SM is neutral under U(1)$'$. This is the scenario
 we will explore here, as the vector nature of the mediator has important 
consequences for the relative fluxes of different annihilation
 products.
 
We assume that the  hidden sector is a singlet under the SM gauge group, while carying a charge under
U(1)$'$. We further assume that $U'(1)$ is spontaneously broken, and restricting the discussion to renormalizable 
couplings, choose a Lagrangian of the following form:
\be
{\cal L}_{ \rm WIMP+V-portal} = {\cal L}_{ \rm WIMP}-
\fr{1}{4}V_{\mu\nu}^2 +\fr{1}{2} m_V^2 V_\mu^2 + 
\kappa V_\nu \partial_\mu F_{\mu\nu}
+{\cal L}_{h'}.
\label{VdF}
\ee
Here we have retained the mixing of $V$ with the photon field strength 
$F_{\mu\nu}$, and combined the residual Higgs$'$ terms in ${\cal L}_{h'}$.
The simplest form of the WIMP Lagrangian is either a Dirac fermion or complex scalar charged under  U(1)$'$, 
\ba
{\cal L}^f_{ \rm WIMP} = \bar \psi (iD_\mu\gamma_\mu - m_\psi) \psi + {\cal L}(\Delta m_\psi)~~~~~~~~{\rm fermionic ~DM},
\\
{\cal L}^b_{ \rm WIMP} = (D_\mu\phi)^* (D_\mu\phi) - m_\phi^2\phi^*\phi+ {\cal L}(\Delta m_\phi)~~~~~~~~~{\rm bosonic ~DM},
\ea
where $D_\mu = \ptl_\mu +ie'V_\mu$ is the usual covariant derivative in terms of the U(1)$'$ gauge coupling $e'$. 
The WIMPs $\ps$ or $\phi$ will be secluded provided $m_V < m_{\ps(\ph)}$, and for the present discussion
 the relevant regime will be $m_V \ll m_{\ps(\ph)}$. 
In general, one can introduce mass terms  ${\cal L}(\Delta m_{\psi(\phi)})$ that lift the degeneracy 
between $\psi$, $\phi$ and their charge-conjugated copies. To be consistent with U(1)$'$ gauge invariance 
the mass splitting has to be proportional to the scale of spontaneus symmetry breaking, $\Delta m_{\psi,\phi} \sim 
v' = m_V/e'$, which in this paper we choose to be small. While we note that 
such a splitting may introduce additional interesting signatures for  direct and indirect 
detection \cite{split1,split2,split3}, we leave this issue aside here and treat 
$\psi$ and $\phi$ respectively as a pure Dirac fermion and a charged scalar.

\subsubsection*{2.1 Relic Abundance} 

\begin{figure}
\centerline{\includegraphics[bb=0 400 500 740, clip=true, width=8cm]{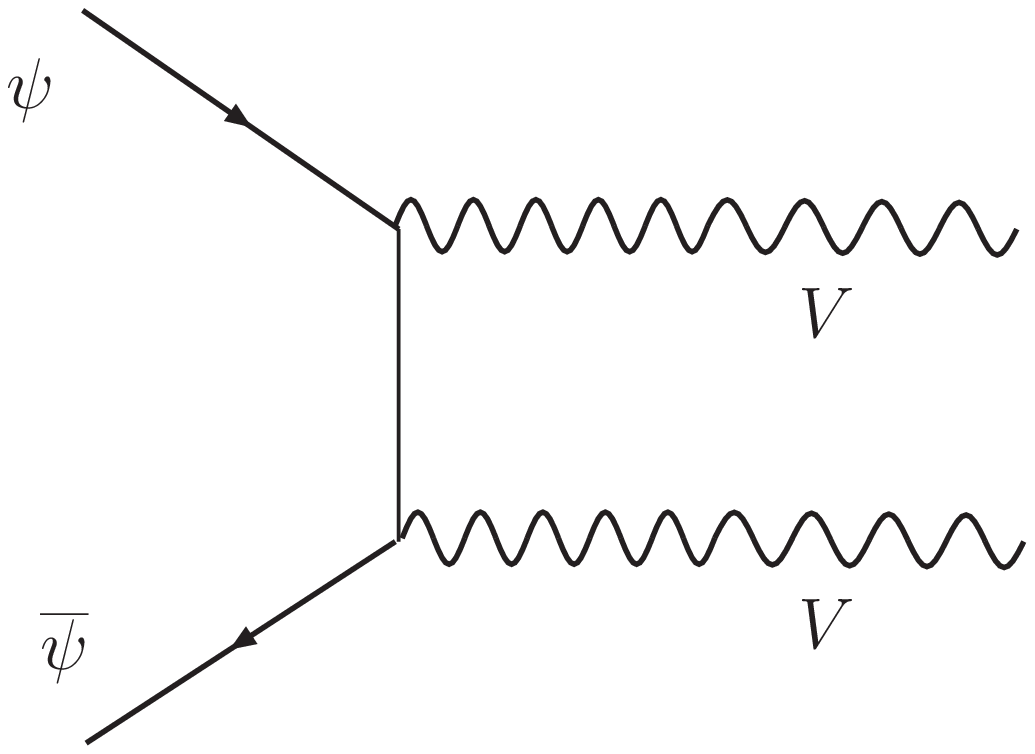}
  \includegraphics[bb=0 485 480 740, clip=true, width=10cm]{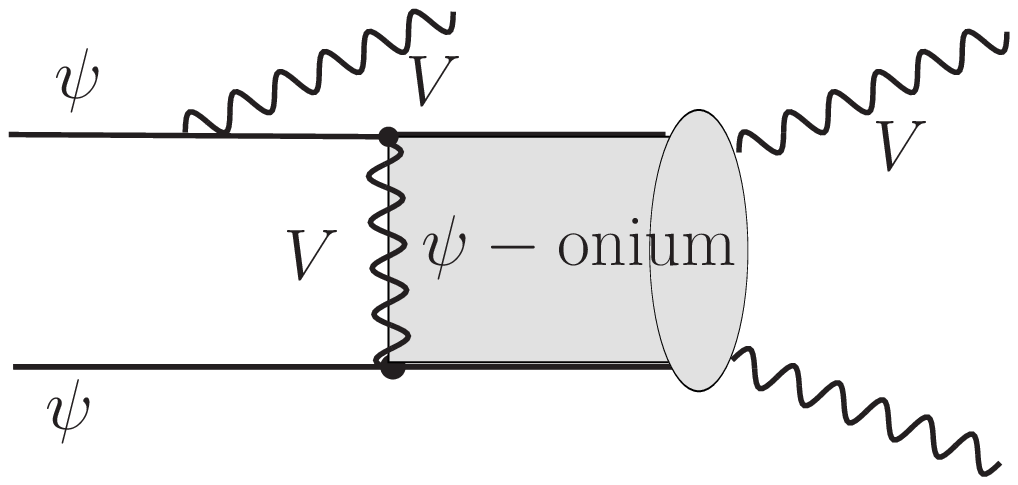}}
\vspace*{-0.6in}
 \caption{\footnotesize  WIMP annihilation: on the left, the direct decay to two metastable on-shell $V$'s, which ultimately decay to SM states; and on
 the right, the decay through the formation of a metastable (para) WIMP-onium state, which occurs with an enhanced rate for non-relativistic WIMPs in
 the galactic halo.}
\label{f1} 
\end{figure}

The dominant annihilation process for determining the relic abundance was discussed in \cite{prv1}.
Provided $m_V<m_\ps$, the two-step process takes the form (see Fig.~\ref{f1}):
\begin{enumerate}
\item[(1)]  $\psi$ + $\bar\psi$ $\to$ on-shell $V+V$,
\item[(2)]  $V$ $\to$ virtual $\gamma, Z$ $\to$ SM states.
\end{enumerate}
The cosmological abundance is determined only by  process (1), and the 
spin-averaged cross section for $\psi^+\psi^-$ annihilation into pairs of (unstable) $V$ bosons is then given 
by
\be
\left. \sigma v \right|^{+-}_{2V}= \fr{\pi (\alpha')^2}{m_\psi^2}
\sqrt{1-\fr{m_V^2}{m_\psi^2}} \stackrel{m_V \ll m_\ps}{\longrightarrow} \fr{\pi (\alpha')^2}{m_\psi^2}, 
\label{2V}
\ee
while for $\phi^+\phi^-$ annihilation the result of (\ref{2V}) should be multiplied by 2.
In order to use (\ref{2V}) in computing the relic cosmological abundance, one has to average 
it over the three different collision types: $+-$, $++$ and $--$, where $+$ and $-$ refer to the U(1)$'$
charge, so that in the absence of a significant primordial asymmetry between 
particles and antiparticles in the WIMP sector,
$\langle \si v \rangle = \fr{1}{2} \left. \sigma v \right|^{+-}_{2V}$. 
Assuming that
$\psi$ (or $\phi$) is the dominant component of dark matter,  and relating the cross section to the
measured  cold dark matter energy density, we obtain:
\be
 \frac{10^{-10} x_f}{\sqrt{g_*(T_f)}
\times  \langle \si v \rangle} \leq \Om_{\rm DM} h^2 \approx 0.1 ~~~\Longrightarrow ~~~ 
\langle \si v \rangle = 2.5\times 10^{-26} {\rm cm}^3 {\rm s}^{-1},
\label{freezeout}
\ee
very close to the conventional order-of-magnitude estimate.
Here $x_f $ is the inverse freeze-out temperature in units of the WIMP mass ($x_f= m_\ps/T_f \sim 20$), and 
$g_*\sim O(100)$ is the number of effective degrees of freedom. Eq.~(\ref{freezeout})
 implies that in the limit of small mixing, $\kappa\ll 1$,
the correct dark matter abundance is achieved if the WIMP masses and couplings satisfy the relation(s)
\ba
\alpha'
\simeq 10^{-2} \times \left(\fr{m_\psi}{ {\rm 270~ GeV}}\right),~~~~~
\alpha'\simeq 10^{-2} \times \left(\fr{m_\phi}{ {\rm 380~ GeV}}\right).
\label{conditions}
\ea
(These formulae also correct a small numerical error in Eq.~(13) of Ref. \cite{prv1}.) 
The conditions (\ref{conditions}) are easily satisfied for a 
rather natural range of $m_{\psi(\phi)}$ and $\alpha'$. The constraints that arise on the
mixing parameter $\ka$, which determines the ultimate decay rate of $V\rightarrow\,\,$SM states, are very mild and $\ka$ can be taken very small
indeed. As discussed in \cite{prv1}, the only constraints one has to impose are 
that the decay of $V$ (and also $h'$) occur before the start of Big Bang nucleosynthesis (BBN). 
Given $m_h' > m_V/2$, only the decays of $V$ are 
sensitive to the mixing:
\be
\Gamma_V \geq {\rm s}^{-1} \quad \Longrightarrow \quad    \kappa^2 \left(\fr{m_V}{10~ {\rm MeV}}\right) \ga 10^{-20}.
\label{BBNkappa}
\ee
If the mediators are light, then $m_V$ is also constrained by BBN 
even if the $V$ bosons remain in thermal equilbrium. An analysis of BBN in the presence of unstable MeV-scale relics 
coupled to the electron-positron plasma \cite{SR} shows that relics with mass
above 4~MeV are consistent with observations, and thus we adopt this value as a lower bound
in our analysis, $m_V \ga 4$~MeV. 
Requiring in addition that $V$ decays remain in thermal equilibrium, i.e. 
$\Gamma_V \geq {\rm Hubble~ Rate}\, [T\simeq 0.05 m_\psi]$ at freeze-out,
would ensure the initial thermal and chemical equilibirium for WIMPs  used in  
the derivation of the abundance formula, leading to a significantly tighter constraint on $\kappa$ \cite{prv1}. However,
this may  be relaxed back to (\ref{BBNkappa}) if some new UV physics ensures proper thermal contact between the dark matter and 
SM sectors at higher temperatures. For the rest of this paper we assume that $\kappa \ll e'/e$, 
so that its precise value will not play a role in calculations of the WIMP abundance and astrophysical signatures 
of WIMP annihilation. To define a fidicual parameter range, we shall mostly consider 
coupling constants $\alpha'$ on the order of the SM gauge couplings at the weak scale, and 
take the dark matter mass to lie in a rather liberal range,  100 GeV $\la m_{\ps(\ph)} \la $ few TeV. 

The recoil signal in direct detection 
depends rather sensitively on all the parameters of the model: $\alpha'$, $m_V$, $\kappa$ and $m_{\psi(\phi)}$.
The relevant WIMP-nucleus scattering cross sections have been calculated previously in \cite{prv1,ptv},
which for $m_V > \mu v$ -- where $\mu$ is the reduced mass and $v$ the relative velocity of the WIMP-nucleus system -- is mediated by the 
electromagnetic charge radius of the WIMP: $r_c^2 = 6 \kappa e'/(em_V^2)$.
The cross section per nucleon in this limit is
\be
\label{sigma_el}
\sigma = \left(\fr{Z}{A}\right)^2\times \fr{16\pi\kappa^2\alpha\alpha'm_p^2}{m_V^4}.
\ee
For $m_V$ of electroweak scale, small values of the mixing parameter $\kappa$ 
strongly suppress the scattering cross section relative to its `natural' weak-scale value, 
implying that the WIMP is indeed `secluded'. However, the limit of small $m_V$ considered in this paper may
entail significant direct-detection sensitivity to the parameter space of the model owing to the $m_V^{-4}$ scaling of
(\ref{sigma_el}). 
The current experimental sensitivity to the WIMP-nucleon cross section, $\sigma \sim 10^{-43}$ cm$^2$,
in the most favorable range of $m_{\psi(\phi)}\sim 100$ GeV,
probes the following combination of masses and couplings,
\be
\label{dirdet}
\kappa^2 \times \left(\fr{\alpha'}{10^{-2}} \right)
\left( \fr{100~ {\rm MeV}}{m_V}\right)^4 \sim {\rm few} \times 10^{-17},
\ee
which is quite close to the lower bound on $\kappa$ in Eq.~(\ref{BBNkappa}).  
Note that for $m_V < \mu v$ this formula needs to be modified to  correct for the dependence
on the energy of the scattering particles. 
Although (\ref{dirdet}) suggests impressive sensitivity to $\kappa$, it should be emphasized that
even a modest splitting $\Delta m_{\psi(\phi)} \sim O({\rm MeV})$ in the dark sector removes 
this completely as $r_c$ is zero for the Majorana fermions or real scalars \cite{split3},
while off-diagonal transitions are kinematically forbidden for $v\sim 10^{-3}$. 
As a consequence, much larger values of $\kappa$ can still be accomodated for small $m_V$.

\subsubsection*{2.2 Enhanced annihilation inside the halo}

Before we present the annihilation rates in  secluded WIMP models, we would like to briefly comment on 
the generic mechanisms for enhancing the annihilation rate in the halo relative to the freeze-out value of 
$\sim3\times 10^{-26}{\rm cm}^3{\rm s}^{-1}$. We introduce a dimensionless parameter, 
\be
{\cal N} \equiv \frac{\langle \si v \rangle_{\rm halo}}{\langle \si v \rangle_{\rm freeze-out}},
\ee
that in the Born approximation is expected to be of order one if the annihilations are predominantly $s$-wave, 
and much less than one for $p$-wave annihilation. This expectation is based on 
the fact that the halo velocities of WIMPs are much smaller than the corresponding velocities at the early cosmological 
 freeze-out epoch. For the purposes of this paper it is sufficient to assume an approximately Maxwellian velocity distribution 
inside the halo with `temperature' $T_h$ and introduce a parameter $x_h$,
\be
x_h = \fr{m_\psi}{T_h} \sim 3\times 10^6,
\ee
that corresponds to the choice of an r.m.s. WIMP velocity of $v_h \sim 200$ km/s. 
Below we list the important exceptions that allow for ${\cal N} \gg 1$:
\begin{itemize}

\item   A: {\em Long-range attractive interaction between WIMPs.} 
In this case, the enhancement is given by the familiar Sommerfeld factor that accounts for the modification of the
scattering wavefunction, relative to the leading-order Born approximation, in the presence of an attractive force. 
We will assume: (i) that the annihilation is $s$-wave; (ii)  an exaggerated hierarchy of scales in Eq.~(\ref{condition1}), namely  $m_V^{-1} \gg a_B$ and $m_V \ll \De p_\ps$
where $a_B\sim (\alpha' m_\psi/2)^{-1}$ is the `Bohr radius' and $\De p_\ps\sim m_\ps v_h$ is the relative WIMP momentum. This then implies as a consequence
 that $\al'/v_h \gg 1$, and  the enhancement factor is
\ba
\label{Sommerfeld}
{\cal N} = 
\left\langle \fr{\pi \alpha'}{v} \right \rangle_h. 
\ea
For simplicity we have disregarded any modifications of the freeze-out rate, assuming that $\alpha'/v_f \ll 1$.  

\item   B: {\em Accidental near-threshold resonances.} Very narrow, near-threshold, 
resonances can enhance the halo annihilation rate quite significantly. 
Such resonances may be due to the accidental presence in the particle spectrum of a mediator with the 
quantum numbers of the WIMP pair and a mass $m_V=2m_\psi$. It is more natural, however, 
to expect a resonance in the form of a bound state of two WIMPs, which is possible if there is an
attractive force acting in the WIMP sector. 
It is not difficult to estimate the maximum possible enhancement in this case. It is well-known that if the [thermal] spread in 
energy of interacting non-relativistic particles is much larger than the resonance width, the corresponding cross section can be written 
as a delta-function in energy, 
\be
\sigma(v) = \fr{4\pi^2}{m_\psi E}
\Gamma_{\rm res} \delta(2E-E_R),
\ee
where $\Gamma_{\rm res}$ is a combination of the total, incoming, and outgoing widths as well as a multiplicity factor
$g$, $\Gamma_{\rm res}= g \Gamma_{\rm in}\Gamma_{\rm out}/\Gamma_{\rm tot}$, and $E$ is the WIMP energy in 
the c.o.m. frame. If the resonant energy is 
within the energy distribution, the ratio of the two thermal rates is maximized and can be as large as 
\be
{\cal N}^{\rm max} = \left(\fr{x_h}{x_f}\right)^{3/2} \sim  O(10^6).
\ee
Should all the requisite conditions be satisfied, a near-threshold resonance may indeed 
result in such a dramatic enhancement although in general the impact on the freeze-out rate, that we have ignored here, may not
be negligible \cite{gs}. However, from the model-building perspective the probability of a narrow resonance right
at the $E=0$ threshold appears 
rather low. Indeed, for $m_\psi$ in the TeV range the maximum resonant enhancement 
occurs when the resonant level is within $\sim 1 $ MeV of the threshold, which would be a rather special point in 
the $\{m_\psi,m_V,\alpha'\}$ parameter space. To have a bound state in the atractive Yukawa potential, 
one would typically expect the interaction range $m_V^{-1}$ to be on the order of or larger than 
$(\alpha' m_\psi)^{-1}$, and the heavier $m_V$ and $m_\psi$ are, the more severe the fine-tuning will become. 

\item   C: {\em New annihilation channels.} While  the two previous mechanisms have already been discussed at some length in 
the literature \cite{Som1,Som2}, a third possibility - namely the enhancement due to new annihilation channels - has not been widely 
noticed.\footnote{Recombination through a Coulomb-like force for a sub-dominant component of dark matter was considered in \cite{bkls}.} 
Specifically,  we refer to the new recombination process (see Fig.~1), 
\be
\label{recomb}
  {\rm recombination:} \;\; \ps + \ps \rightarrow (\ps-{\rm onium}) + V,
\ee
(and similarly for $\ph$)
which is kinematically open even in the limit  $E_\psi \to 0$  if the condition (\ref{condition2}) is satisfied. 
The subsequent fate of the $\ps$-onium state is very different within the early Universe during freeze-out as compared to the 
galactic environment. In the halo, every $\ps$-onium that is formed via the process (\ref{recomb}) decays further 
to two or three $V$-bosons.  During freeze-out, however, the annihilation rate of $\ps$-onium into $V$'s is strongly
inhibited by thermal break-up, $\psi$-onium + V $\rightarrow$ $2\psi$. One can easily show that in the latter case
the efficiency of annihilation, ${\rm Br} = \Gamma_{\rm annih}/(\Gamma_{\rm annih} + \Gamma_{\rm break-up})$, is much smaller than one. 
Thus, effectively only when the temperature drops below the binding energy does the process (\ref{recomb}) serve as 
a new annihilation channel and, as we are going to see shortly, 
indeed dominate the annihilation rate in the galactic halo. 
\end{itemize}

When the recombination process (\ref{recomb}) is kinematically allowed according to (\ref{condition2}) -- 
which for $m_\ps\sim 500$~GeV and $\al'$ fixed according to (\ref{conditions}) requires $m_V\la 50$~MeV --
the rate can be computed by generalizing the 
corresponding calculation for positronium to a finite vector mass $m_V$. Retaining only the direct recombination 
to the ground state of WIMP-onium, we arrive at a recombination rate that is independent 
of the spin of the annihilating particles. Assuming once again that $m_V \ll \De p_{\ps(\ph)} \sim m_{\ps(\ph)} v_h$, which is
the range of parameters relevant here, we find
\be
 \left. \si v \right|^{+-}_{\rm rec} = \frac{2^{10} \pi^2(\al')^2}{3 \exp(4) m_{\ps(\ph)}^2}
\left(\frac{v_V(3 - v_V^2)}{2}\right)\left\langle \fr{\alpha' }{v} \right \rangle_h, 
 \label{rec}
 \ee
where $v_V = [1 - 4m_V/(\alpha'm_{\psi(\ph)})^2]^{1/2}$ is the velocity of the emitted $V$ boson,
which we assume is not significantly different from 1. 
Similar WIMP recombination processes were considered previously is Refs.~\cite{split3,BKP}.
The $\ps(\ph)$-onium state will eventually decay to either $2V$ or $3V$,  
with branching ratios of 25 and 75\% for the fermionic WIMPs, and to $2V$ for bosonic dark matter. 
Since $m_{\ps(\ph)}$ is large, the decay rates are rather fast, 
e.g. $\Ga^{\ps}_{2V} \sim (\al')^5m_\ps$, followed by 
the subsequent much slower decays of $V \rightarrow\;$SM states
with lifetimes controlled by the small mixing parameter $\ka^2$.

The crucial observation is that this recombinant annihilation 
process is significantly enhanced in the galactic halo relative to the direct decay to $2V$ which
dominates at freeze-out. Besides the expected $\alpha'/v$ factor, there is an additional 
well-known enhancement by the large numerical coefficient in (\ref{rec}).
Comparing (\ref{rec}) to (\ref{2V}), we find the enhancement factors
for the cases of fermionic and bosonic dark matter:
\be
\label{Nrec}
{\cal N}^{\psi} \simeq 20 \left\langle\frac{\al'}{v}\right\rangle_h, ~~ 
{\cal N}^{\phi} \simeq 10 \left\langle\frac{\al'}{v}\right\rangle_h .
\ee
For $ \alpha \la \al' \la \alpha_w$ this implies a large enhancement factor of ${\cal O}(100)$, 
  as $\langle v^{-1} \rangle_h \simeq  (2x_h/\pi)^{1/2} \sim 1.5 \times 10^{-3} $.
  Moreover, this enhancement will occur  as long as the conditions (\ref{condition1}) and (\ref{condition2}) are satisfied, and this does not require
  any tuning of resonant levels. The 
inclusion of recombination to $2s$ WIMP-onium states, which is allowed if $16 m_V < (\alpha')^2m_\psi$, would 
further increase this number by $\sim 20\%$. 
  If the emission of $V$ bosons in recombination is not kinematically allowed, i.e. if $m_V > E_{\rm bind}$, 
but the conditions in (\ref{condition1}) are still satisfied, the enhancement of e.g. the $\psi + \psi \to 2V$ rate is given 
simply by Eq.~(\ref{Sommerfeld}), which for the same value of $\alpha'$ is a factor of $\sim$7(3) times smaller 
than ${\cal N}^{\psi(\phi)}$  in (\ref{Nrec}) for fermionic(bosonic) WIMPs. 

The final step in the decay chain is the decay of the $V$ bosons resulting from 
WIMP annihilation (possibly within a WIMP-onium bound state).  Depending on the value of $m_V$, one may have 
different SM final states. In a generic secluded regime, $m_V \le m_{\ps(ph)}$, 
such final states may involve leptons, both light and heavy quarks, and $W$ pairs. Consequently, 
one should expect a non-negligible fraction of anti-protons being created in the process of hadronization of 
the decay products. However, in the domain of parameters where the halo 
annihilation rates are most significantly enhanced, $m_V < E_{\rm bind}$, only decays to light charged mesons and electron and 
muon pairs are kinematically possible, as $E_{\rm bind} \la 1$ GeV even for multi-TeV WIMPs. 
As a consequence, all SM decay products will be highly boosted, and thus can provide 
attractive targets for indirect detection as $\gamma$-rays and positrons with energies
of ${\cal O}(100)$~GeV have a rather limited scope for production through conventional astrophysical sources. 
The final parameters of interest in this scenario are the relative branching fractions for the 
decay of $V$, which occurs through  mixing with off-shell photons.
As it turns out, the vector nature of the mediator in this case 
leads to a dominant leptonic branching fraction (e.g. $V\rightarrow \gamma^* \rightarrow e^+e^-$),
as direct decays to $2\gamma$ are forbidden and the lowest order photonic decay rate is $V\rightarrow 3\gamma$
proceeding via a loop diagram. Direct decays of $V$'s to neutrinos are also highly inhibited. 
Since all light charged mesons and muons eventually cascade down to electrons, positrons and neutrinos, 
we conclude that on average between two-to-three positrons are produced for each WIMP annihilation, 
while the yield of $\gamma$-quanta, resulting from internal bremsstrahlung  in $V$-decays, is expected 
to be suppressed by an $O(\alpha/\pi)$ factor\footnote{We note in passing that for secluded 
WIMP models with scalar mediators in the $O(100$ MeV) range, the branching fraction of $\gamma$  
quanta in the decay products will be significantly enhanced and can indeed dominate over
other decay channels. }. 
The dominant leptonic branching fraction has interesting observational 
consequences that we will explore in the next section.

\subsection*{3. Observational Signatures}

The dominant observational signals for DM models of this type comprise an excess of high-energy $\gamma$-rays and 
positrons from annihilation. 
Since astrophysical sources generally fall as power-laws, the higher energy end of the spectrum, 
$E\sim$ few 100 GeV, is the region with the best signal/background ratio.
It also means that for the class of secluded models considered here, the $\gamma$-ray production
rate (at least locally) will be of less importance than the positron rate, 
because of the $\sim \alpha/\pi$ suppression mentioned in the previous subsection. 

Turning to the observable flux, it is worth recalling that the signatures for both 
$\gamma$-rays and positrons are highly dependent on details of
the galactic halo, but in different ways. 
The observed $\gamma$-flux is proportional to the line-of-sight integral, 
$\int_{\rm los} dl \langle \si_\gamma v\rangle_h \rh(l)^2$, where 
$\rh(l)$ is the dark matter  density along the line of sight. The flux thus
samples the entire halo distribution, and is enhanced in directions where the density is 
largest, e.g. the galactic center. It is therefore
sensitive to the fact that the inner halo profile is still not well 
constrained by simulations. Recently, several high energy sources believed to be of
astrophysical origin have been detected in the direction of the galactic 
center by HESS \cite{hess} and EGRET \cite{egret}, which unfortunately enhances the 
background in directions where the signal should be at its peak. 
In contrast, the observed positron flux, at least for energies near the 
peak of the decay spectrum, is a more `local' signal.
The observed flux is affected significantly by energy degradation 
through various loss mechanisms, such as synchrotron
radiation and inverse Compton scattering. The flux can be represented in the form,
\be
 \Ph_{e^+(E)} =  \langle \si v \rangle_{e^+} 
\left(\frac{\rh_0}{m_\ps}\right)^2 \int_{E_{\rm in}>E} dE_{\rm in} 
\frac{dN_{e^+}}{dE}(E_{\rm in})F_P(E,E_{\rm in},\lambda_D),
  \label{pos_flux}
\ee
where $\rh_0$ is the constant normalization of the DM density in the local region, and 
$dN_{e^+}/dE$ is the annihilation spectrum. For the $2V$ final state, the emitted
$V$ bosons are monoenergetic, which creates a flat spectrum of 
 $e^+e^-$ pairs in the `halo restframe',  $dN_{e^+}/dE = N_0 \Theta(m_\ps-E)$. 
The propagation function $F_P$ depends on the diffusion 
length $\lambda_D(E,E_{\rm in})$, which determines the energy degradation from the 
injection energy $E_{\rm in}$ to the observed energy $E$, 
and in general many other details of the 
halo distribution (see e.g. \cite{dldfs}). However, due to this 
energy degradation, if we focus only on the high-energy part of the spectrum 
where $E \sim m_\ps$ this effectively samples just the local 
neighbourhood where the dark matter density should be roughly constant. 
For $E\sim m_\ps\sim 500$~GeV, the diffusion distance is probably 
at most 1~kpc.  In this approximation,  $F$ is obtained 
by integrating just over a local sphere and thus, while a full solution 
of the diffusion equation \cite{prop} will introduce 
some sensitivity to the halo profile, the high-energy part
of the positron flux is primarily sensitive only to the local DM number density, 
as is also the case for direct-detection. Although $\rh_0$ is itself quite uncertain, e.g. due 
to halo substructure, this effectively limits the DM-related 
uncertainties to the combined combination of $\rh_0^2 \langle \si v \rangle$.

 \begin{figure}
\centerline{\includegraphics[width=8.5cm]{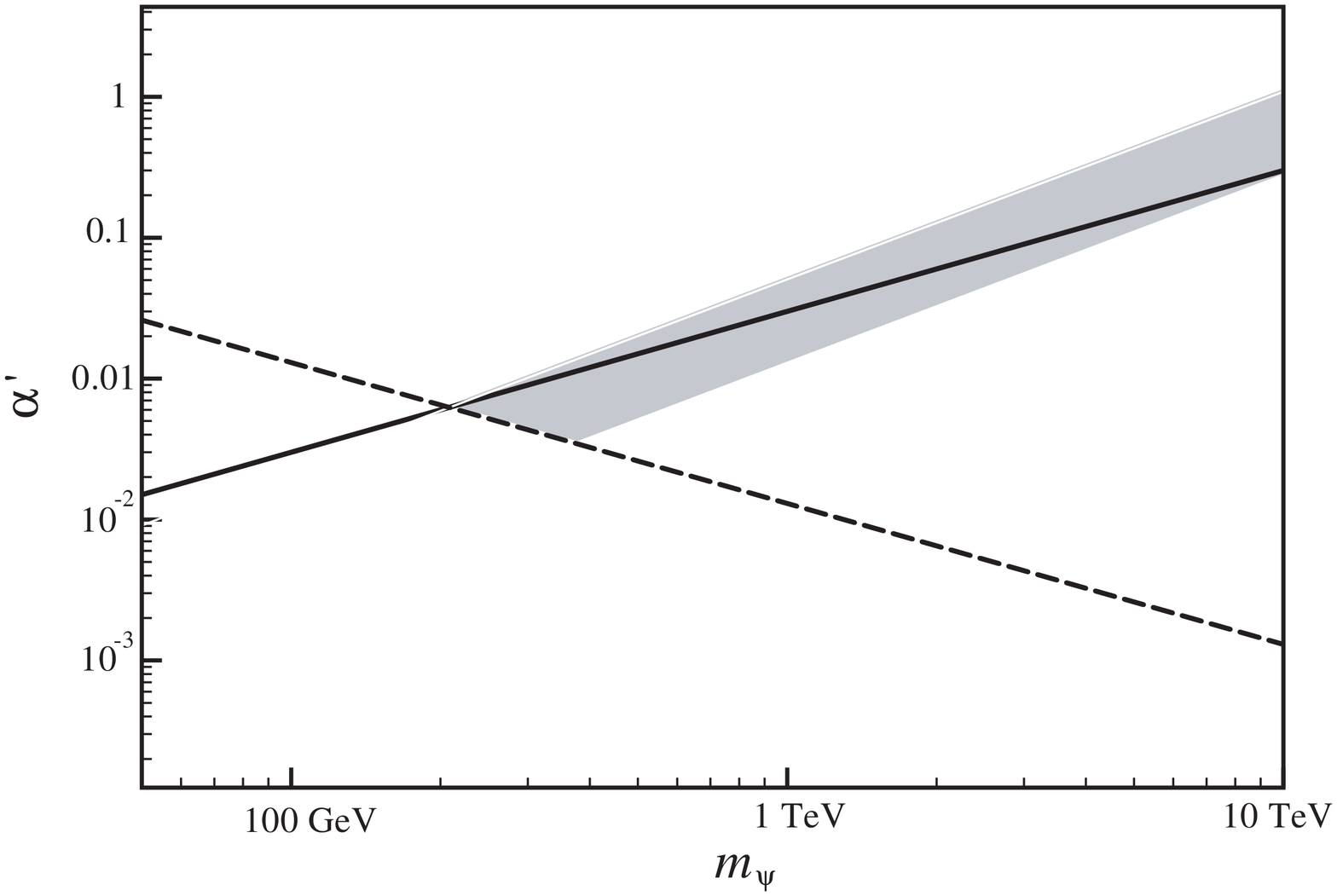}\includegraphics[width=8.5cm]{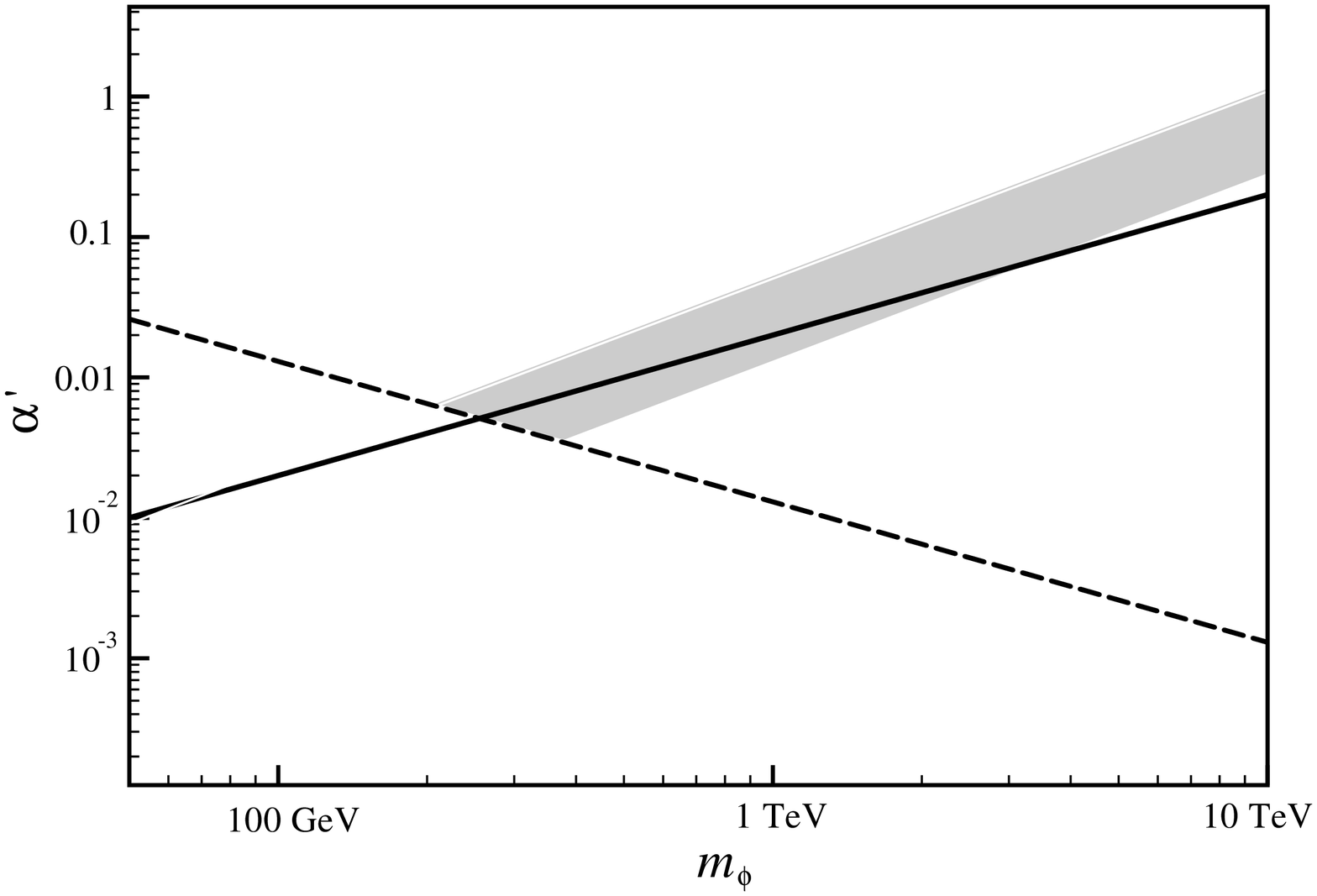}}
 \caption{\footnotesize  
The $m_{\psi(\phi)}$-$\alpha'$  parameter space for fermionic (left) and bosonic (right) secluded WIMPs,
showing the cosmological abundance constraint (solid line), the region where radiative recombination into WIMP-onium is allowed (to the right of the 
dashed line), and a grey band where, limited by two extreme models of $e^+$ propagation, the positron excess would be above background for energies above
10~GeV. }
\label{f2} 
\end{figure}

For secluded DM models with a vector mediator, we showed in the 
previous section that the dominant primary product of
annihilation will be $e^+e^-$ pairs, rather than $\gamma$-rays. 
While $\gamma$'s dont suffer the same energy degradation and so can have a larger observable 
flux due to annihilation throughout the halo, particularly when combined with secondary $\gamma$-production 
through bremsstrahlung and synchrotron emission, we will focus now on 
the observability of the positron component. As discussed earlier, the recent preliminary announcement 
from the PAMELA collaboration, apparently supporting the existing suggestive results from HEAT and AMS-01 of an excess positron 
fraction above 10~GeV, is difficult to reconcile with conventional WIMP models as it requires an annihilation 
cross-section well above that expected for a conventional thermal WIMP, i.e. $\langle \si v\rangle \sim 3\times 10^{-26}$cm$^3$s$^{-1}$. 
Thus, to explain the data an astrophysical `boost factor'  
$\langle \rh^2\rangle/\langle \rh\rangle^2\gg 1$ is often postulated reflecting the 
possibility of a local enhancement in the number density due to substructure in the halo. 

Various analyses \cite{olderDM, th-speculation} indicate that the boost factor required to a produce a 
positron excess in the HEAT range, i.e. $n_{e^+}/(n_{e^-}+n_{e^+})\sim 0.1$
above 10-20~GeV, is roughly:
\be
 \langle \si v \rangle_{e^+} \sim  K 
\left(\frac{m_\ps}{500\,{\rm GeV}}\right)^2  \langle \si v \rangle_{\rm freeze-out}\;\;\;\;\;{\rm where} \;\;\; K \sim \frac{{\cal O}(100)}{N_*},
 \ee
 where $N_*$ is a multiplicity factor counting the number of positrons produced for each annihilation. In the present case, given
 the various decay channels through ortho- and para- states of WIMP-onium, we have $N_* \sim 2$.
 Our primary observation is that this boost-factor, although somewhat uncertain, 
appears to be of precisely the order of magnitude 
  achieved through recombination of WIMPs
 in the galactic halo as discussed in the previous section. Thus, in such scenarios, 
one would expect an observationally detectable positron
 excess even in the absence of astrophysical boost factors.
 
 To illustrate the interesting region of parameter space, we plot in Fig.~2 the $m_{\psi(\phi)}-\alpha'$ plane taking
 the extreme value $m_V \simeq 4$ MeV to maximize the allowed region  where annihilation occurs via radiative recombination. This 
 line intersects the constraint on the primordial abundance at around 200 GeV, and the region to the right is allowed. The grey band 
 illustrates a representative range, allowing for unknown parameters in the propagation of positrons in the local region, that would produce
 an enhanced cross-section consistent with the observed excess seen by HEAT and AMS-1. The two extreme propagation models were taken
 from \cite{prop}.  This band is almost parallel to the cosmological abundance line, but the enhancement factors ${\cal N}^{\psi(\phi)}$ lead to
an overlap region which covers a large mass range due to the uncertainties in positron diffusion. Indeed, we see that there is  broad 
consistency between the cosmological constraint and the apprent level of the positron excess for $200$ GeV $\la m_\psi \la$ 10 TeV, 
and $200$ GeV $\la m_\phi \la$ 4 TeV, given a value for $m_V$ consistent with (\ref{condition2}). It is also worth noting that if
$m_V$ lies somewhat above the recombination range (\ref{condition2}), the Sommerfeld enhancement factor (\ref{Sommerfeld}) still
applies, and so the grey band can be continued to the left of the dashed line in Fig.~2, but shifted upward according to the reduction in the
annihilation rate.

\subsection*{4. Discussion}

In this note, we have presented a simple scenario within which the 
annihilation cross-section of a thermal WIMP may be significantly
enhanced in the galactic halo with respect to its standard picobarn rate at freeze-out. 
For most WIMPs, the astrophysically relevant annihilation rate $\langle \sigma v\rangle$ 
is equal to or smaller than the corresponding rate at freeze-out. However,  for some models this general rule
is violated, and  we discussed three generic mechanisms that could boost the halo annihilation signal: (A)
(relatively) long-range interactions in the WIMP sector; (B) accidental near-threshold resonances; and (C) new annihilation channels.
While (A) and (B) have been considered previously, we focused on the third possibility where the mechanism relies on the possibility
of annihilation proceeding through recombination, via a metastable WIMP-onium bound state, and thus requires the presence of a 
light mediator and thus long-range forces in the dark sector. We observed that the recombinant rate is naturally enhanced by a factor of
${\cal O}(100)$, while at the same time such models characteristically allow for a significant decoupling, or seclusion, of the dark sector with a small
cross-section for scattering on nuclei. The vector nature of
the mediator tends to enhance the leptonic decay fraction relative to $\gamma$'s, which is particularly intriguing given 
recent speculation about a high-energy positron excess. While the excess observed by HEAT, AMS-01, and now 
in preliminary form by PAMELA, may find a conventional astrophysical explanation, it is interesting that a fairly
generic class of WIMP models, namely secluded models with light mediators, can naturally have an enhanced
annihilation cross-section in the halo that may have additonal consequences that would be interesting to explore
further. We will conclude by mentioning several other  aspects of secluded WIMP models.

\begin{itemize}

\item {\em Implications for BBN:} It is well known that the annihilation 
of standard neutralino-like WIMPs during BBN does not affect the light elemental abundances 
unless the neutralino mass is under $\sim 80$ GeV \cite{Jedamzik}. 
However, the enhancement of the annihilation cross section resulting from a long-range interaction in 
the dark sector will change this conclusion as the number of WIMP annihilations after freeze-out 
acquires a different (milder) scaling with temperature. It is customary 
to characterize the impact of decaying or annihilating particles on the light 
elemental abundances by the quantity $\xi$ -- the energy release in 
GeV normalized by the thermal photon number density. 
Assuming that all the energy from annihilation is released in the form of 
electromagnetic radiation, which is indeed the case for the models considered here,
and following the earlier calculation in \cite{BKP}, we estimate that the value of $\xi$ due to annihilation 
at $T\sim 0.1$ keV can be as large as a few$\times 10^{-13}$, which will have an impact on the 
late-time generation of $^6$Li at an observationally interesting level of $^6$Li/H $\sim 10^{-11}$.

\item {\em Signatures in the visible sector}:
Among the allowed `portals', 
i.e. renormalizable couplings of the mediator to the SM: mixing of a U(1)$'$
vector with the photon; coupling of a singlet scalar to the Higgs sector; 
or perhaps the right-handed neutrino, existing theoretical prejudice on the mass
scale suggests that only the U(1)$'$ can naturally be much lighter than the weak scale. 
When the scale of the mediator mass is much smaller than the weak scale
but the coupling between the secluded and visible sectors is not too small, a new experimental signature
 may become available \cite{prv1,Fayet}. For mixing angles 
of order $\kappa \sim O(10^{-3})$ between the visible and secluded U(1)$'$ sector, as would naturally arise from
integrating out heavy states charged under both hypercharge and the secluded U(1)$'$,
the presence of $O($1-100~MeV)-scale mediators can create small but 
potentially detectable deviations from the SM in rare decays, the muon $(g-2)$ anomaly, and 
precison QED tests \cite{Posp}.

\item {\em Implications for direct detection and rare isotope searches}: 
As noted in Sect.~2, the relatively long range force mediated by 
$V$ bosons  leads to a considerable enhancement  of the direct detection rate, thus limiting the level to which the
dark sector can be secluded. Its worth pointing out that if this effect is  
counterbalanced by the appropriate splitting $\Delta m_{\psi(\phi)}$ of the WIMP states, it may lead to distinct signatures
for the `inelastic' scattering of dark matter \cite{split1}. At the same time, if $m_V^{-1}$ is much larger than the typical 
nuclear size of a few fm,  sizable values for $(e'\kappa)$ could possibly lead to the binding of 
WIMPs with heavy nuclei due  to $V$-exchange, resulting in recombination signatures 
in direct detection and the presence of  stable anomalously heavy nuclei \cite{split3}. However, all
inelastic scattering possibilities are quite sensitive to the parameters of the model.

\end{itemize}

\noindent{\bf Note added:} While this paper was being finalized, two preprints appeared \cite{new} which also discuss the 
relevance of a long-range interaction in the dark matter sector, and overlap with some parts of the present paper.

\subsection*{Acknowledgements}

We would like to thank Misha Voloshin for numerous illuminating discussions on the 
subjects pertinent to this paper. 
The work of MP and AR was supported in part
by NSERC, Canada. Research at the Perimeter Institute
is also supported in part by the Government of Canada through NSERC and by the Province
of Ontario through MEDT. The work of MV is supported in part by the DOE grant DE-FG02-94ER-40823.

\end{document}